\documentclass[12pt]{iopart}
\begin{document}
\title[Transition Matrix, Poisson bracket for gravisolitons]{Transition Matrix, Poisson bracket for gravisolitons in the dressing formalism}
\author{P Kordas}
\address{35 Square Marie-Louise\\1000 Bruxelles\\Belgium}
\ead{panayiotis.kordas@physics.org}

\begin{abstract}
The Hamiltonian methods of the theory of solitons are applied to gravisolitons in the dressing formalism. The Poisson bracket for the Lie-algebra valued one-form $A(\varsigma, \eta, \gamma)=\Psi_{,\gamma}\Psi^{-1}$, for gravisolitons in the dressing formalism, for a specific background solution, is defined and computed, agreeing with results previously obtained. A transition matrix ${\cal T}=A(\varsigma, \eta, \gamma) A^{-1}(\eta, \xi, -\gamma)$ for $A$ is defined relating $A$ at ingoing and outgoing light cones. It is proved that it satisfies equations familiar from integrable pde's with the role of time played by the null coordinate $\eta$. This is a new result mathematically, since there has not been a transition matrix for $A$ in the litterature, while physically it presents the possibility of obtaining integrals of motion (for appropriate boundary conditions), from the trace of the  derivative with respect to the null coordinate $\eta$, of ${\cal T}$, in terms of classical relativity connections, since $A(\gamma = \pm 1)$ can be expressed in a simple way in terms of the classical Christoffel symbols. This may prove of use upon quantization since connections are fundamental variables of quantum gravity.  The roles of $\eta$ and $\varsigma $ may be reversed to obtain integrals of motion for $\varsigma$, thus $\varsigma$ playing the role of time. This ties well with the two-time interpretation and approach already established before.
\end{abstract}
\submitto{JPA}

\section{Introduction}
\label{sec:Intro}
 The general relativistic equations in the presence of two commuting Killing vectors have received a great amount of interest over the years. The first important contribution, relevant to the considerations here, was the proof in \cite{Geroch} that there exists an infinite hierarchy of solutions which can be mapped to one another via transformations of what we now call the Geroch group. 
Later it was shown \cite{BelinskyZakharov1, Maison, Neugebauer, HauserErnst, Harrison, HKX} by different authors and in differing approaches, that the field equations (\ref{eq:field equations g}) are integrable (in the sense understood in the inverse scattering field) and a variety of solutions was obtained and analyzed. These results are presented and reviewed in \cite{BelinskyVerdaguer, KordasReview}. Relatively more recently techniques of quantum inverse scattering were used\cite{KorotkinNicolai} to quantize appropriate functions of the gravitational field appearing in (\ref{eq:field equations g}). This quantization however was incomplete due to not sufficiently developed representation theory of SL(2, R). The emphasis was on the metric with the well-known problem that the quantum operator corresponding to the metric is not known.  The main motivation of the present paper is to present an approach for future work in terms of connections which are appropriate variables for quantum Gravity. A small step in this direction is the current paper which applies some of the ideas in \cite{KorotkinNicolai, SamtlebenThesis} to the dressing formalism for solitons as developped in \cite{BelinskyZakharov1, BelinskyZakharov2}. 

 The organization of the present work is as follows. In section (\ref{sec:BackgroundForEinsteinSEquationsInTheBelinskyZakharovFormalism}) we introduce the equations and formalism of the solitonic technique. Further in section (\ref{sec: spectral current}) the `spectral parameter current'\cite{KorotkinNicolai} is defined and the form of the fundamental Poisson bracket applicable to the case considered is presented. A transition matrix for the `spectral parameter current' $A$ relating $A$ at points on ingoing and outgoing null coordinates is defined and it is proved that it obeys equations similar to the ones of other integrable pde's, a result absent from the literature thus far. This presents the possibility of obtaining integrals of motion, from the trace of the derivative of the transition matrix w.r.t. the null coordinate $\eta$ for appropriate boundary conditions (ie. the role of time is played by the null coordinate $\eta$, a fact which ties well with the `two-time' approach and interpretation\cite{KorotkinNicolai}; the role of time can be played by $\varsigma$ with a different definition of transition matrix thus providing integrals of motion for $\varsigma$), in terms of classical connections of General Relativity. These may be appropriate for QG as they are the fundamental variables of the classical theory and connections are fundamental also in Quantum Gravity \cite{Rovelli, Ashtekar}.  The fundamental Poisson bracket and the transition matrix characterize a wide variety of integrable systems \cite{FaddeevTakhtajan, KorotkinNicolai}, and play a crucial role in quantizing such systems \cite{Korepin}. We conclude in Section (\ref{sec:Conclusion}) with a discussion.

\section{Einstein's equations with two commuting Killing vectors in the Belinsky-Zakharov formalism}
\label{sec:BackgroundForEinsteinSEquationsInTheBelinskyZakharovFormalism}
The metric in the presence of two commuting Killing vectors, and assuming the existence of 2-surfaces orthogonal to the group orbits, is given by \cite{ExactSolutionsBook, BelinskyZakharov1, BelinskyVerdaguer}:
\begin{equation}
\label{eq:metric}
	ds^{2} = f(t,z)(dz^{2}-dt^{2})+g_{ab}(t,z)dx^{a}dx^{b},
\end{equation}
where $g_{ab}(t,z)$ real and symmetric tensor, $a,b= 1, 2$.

Einstein's equations corresponding to this metric are, (in null coordinates $ \varsigma = \frac{1}{2}(z+t)$, $\eta = \frac{1}{2}(z-t)$)
\begin{eqnarray}
\label{eq:field equations g}
(\alpha g_{,\varsigma}g^{-1})_{,\eta}+(\alpha g_{,\eta}g^{-1})_{,\varsigma}=0, \\
\label{eq:field equations f1}
(\ln f)_{,\varsigma}(\ln \alpha)_{,\varsigma} = (\ln \alpha)_{,\varsigma\varsigma}+\frac{1}{4\alpha^{2}} \mbox{Tr} A^{2}(1),\\
\label{eq:field equations f2}
(\ln f)_{,\eta}(\ln \alpha)_{,\eta} = (\ln \alpha)_{,\eta\eta}+\frac{1}{4\alpha^{2}} \mbox{Tr} A^{2}(-1),
\end{eqnarray}
where 
	\begin{equation}
	\label{eq:A(+-1)}
	A(1) = \alpha g_{,\varsigma}g^{-1}, \,\,\,\,\,\,\,\, A(-1) = \alpha g_{,\eta}g^{-1}, 
	\end{equation}
	$ \det g = \alpha^{2} $ and the reason for the labelling $ A(\pm 1) $ will become apparent in the following. Further from the above equations it follows that $ \alpha $ satisfies 
	\begin{equation} \alpha_{,\varsigma\eta}=0. \label{eq:alphaequation}\end{equation}
Belinsky-Zakharov have shown that equation (\ref{eq:field equations g}) is integrable \cite{BelinskyZakharov1,BelinskyVerdaguer,BelinskyZakharov2}. This means the existence of a linear system of differential equations 
\begin{eqnarray}
\label{eq:linear system}
	\frac{d\Psi}{d\varsigma} = \frac{A(1)}{\alpha(1-\gamma)}\Psi , \,\,\,\,\,\,\,\,
	\frac{d\Psi}{d\eta} = \frac{A(-1)}{\alpha(1+\gamma)}\Psi,
\end{eqnarray}
 where we use the notation of \cite{KorotkinNicolai}. Reality is ensured via $\Psi^{*}(\gamma^{*})=\Psi(\gamma)$. The system (\ref{eq:linear system}) has as compatibility conditions equation (\ref{eq:field equations g}) and the zero-curvature condition \cite{FaddeevTakhtajan} associated with this integrable system \cite[p. 15, eq. 1.48]{BelinskyVerdaguer}.
The differentials are given by 
\begin{equation}
 \label{eq:differentials}
 \frac{d}{d\varsigma}=\frac{\partial}{\partial\varsigma}+\frac{\gamma}{2\alpha}   \frac{1+\gamma}{1-\gamma}\frac{\partial}{\partial\gamma}, \,\,\,\,\,\,\,\,
 \frac{d}{d\eta}=\frac{\partial}{\partial\eta}-\frac{\gamma}{2\alpha}  \frac{1-\gamma}{1+\gamma}\frac{\partial}{\partial\gamma},
\end{equation}
where the solution $ \alpha = \varsigma - \eta = t$ of (\ref{eq:alphaequation}) is chosen here and \(\gamma\) is a 'variable' spectral parameter satisfying 
\begin{equation}
 \label{eq:gamma equation}
 \gamma_{,\varsigma}=\frac{\gamma}{2\alpha} \frac{\left(1+\gamma\right)}{\left(1-\gamma\right)},\,\,\,\,\,\,\,
 \gamma_{,\eta}=- \frac{\gamma}{2\alpha} \frac{\left(1-\gamma\right)}{\left(1+\gamma\right)},
\end{equation}
which are solved by 
\begin{equation}
\label{eq:gamma+-}
 \gamma_{\pm}(w;\varsigma,\eta) = \frac{1}{\alpha}\left\{w-\beta \pm \sqrt{(w-\beta)^{2}-\alpha^{2}}\right\} = 1/\gamma_{\mp},
\end{equation}
where $w$ complex constant and $\beta=\varsigma+\eta=z$ is a second solution of (\ref{eq:alphaequation}).

Solutions of equations (\ref{eq:linear system}) can be reproduced from a known background solution according to the following procedure \cite{BelinskyZakharov1, BelinskyZakharov2, BelinskyVerdaguer}: One starts with $\Psi_{0}$ a solution of (\ref{eq:linear system})(corresponding to a background metric $g_{0}$) and form
\begin{equation}
\label{eq: soliton ansatz}
 \Psi = \chi(\varsigma,\eta,\gamma)\, \Psi_{0},
\end{equation}
where 
\begin{equation}
	\label{eq:chi}
	\chi = \mbox{I}+\sum_{k=1}^{N}\frac{R_{k}(\varsigma,\eta)}{\gamma-\gamma_{k}}.
\end{equation}
where I unit matrix, and the poles $\gamma_k$, which have to be solutions of (\ref{eq:gamma equation}) ie. are given by (\ref{eq:gamma+-}) for $w=w_k$, correspond to solitons in the sense understood in the inverse scattering litterature.
Reality is ensured via $\chi^{*}(\gamma^{*})=\chi(\gamma)$.

It is important that the variables $R_{k}$ have no $\gamma$ dependence and are only functions of $\varsigma$ and $\eta$. The poles $\gamma_{k}$ can be interpreted as the null trajectories of perturbations propagating on the background solution and can be thought of as 'gravitational solitons'\cite{BelinskyVerdaguer} although they are not fully analogous to solitons as they are known in other integrable pde's \cite{Kordas PhD, Kordas gravibreather, Gleiser}. 

It can be shown \cite{BelinskyZakharov1, BelinskyVerdaguer} that $\chi$ satisfies the equation 
\begin{eqnarray}
	\label{eq: chi diff equation}
	\frac{d\chi}{d\varsigma} = \frac{1}{\alpha(1-\gamma)}(A(1)\chi(\gamma)-\chi(\gamma) A_{0}(1)) \\
  \frac{d\chi}{d\eta} = \frac{1}{\alpha(1+\gamma)}(A(-1)\chi(\gamma)-\chi(\gamma) A_{0}(-1)),
\end{eqnarray}
where $A_{0}(\pm 1)$ is the variable that corresponds to the background solution.

It is ensured that $g$ is symmetric via
\begin{equation}
\label{eq:chichiT}
g^{-1}\chi(\gamma) = (\chi^{-1}(1/\gamma))^T g_0^{-1}
\end{equation}
The $R_k$ are degenerate matrices and may be expressed as \cite{BelinskyVerdaguer} 
\begin{eqnarray}
\label{eq:Rk}
(R_{k})_{ab} = n_a^{(k)} m_{b}^{(k)} \\
n_a^{(k)} = \sum_{l=1}^{N}\frac{(\Gamma^{-1})_{kl}m_{c}^{(l)}(g_0)_{ca}}{\alpha\gamma_l}\\
\Gamma_{kl} = -\frac{m_c^{(k)}(g_0)_{cb}m_{b}^{(l)}}{\alpha^2(1-\gamma_k\gamma_l)}\\
m^{(k)}_a = m_{0b}^{(k)}[\Psi_0^{-1}(\gamma_k,\varsigma,\eta)]_{ba}
\end{eqnarray}
$m_{0b}^{(k)}$ arbitrary complex vectors.

From \ref{eq:chichiT} it is seen that $\chi^{-1}$ may be expressed as
\begin{equation}
\label{eq:chiinv}
\chi^{-1} = \mbox{I} + \sum_{k=1}^{N}\frac{S_k}{\gamma-\gamma_{k-}}
\end{equation}
The same way that the residues $R_k$ are obtained in \cite{BelinskyVerdaguer} the degenerate matrices $S_k$ may be obtained by expressing \(A_0(\pm 1)/(\alpha(1\mp\gamma))\) from the equations (\ref{eq: chi diff equation}) in terms of $d\chi\chi^{-1}$ and using \(\chi(\gamma_{k-})S_k=0\) which is \(\chi\chi^{-1}=\mbox{I}\) at the poles $\gamma_{k-}$ of $\chi^{-1}$. In the same manner as in \cite{BelinskyVerdaguer} one obtains:
\begin{eqnarray}
\label{eq:Sk}
(S_{k})_{ab} = q_a^{(k)} p_{b}^{(k)} \\
q_{a}^{(k)} = q_{0b}^{(k)}(M_k)_{ba} \\
p_{a}^{(k)} = \sum_{l=1}^{N} \frac{(g_0^{-1})_{ac}q_c^{(k)}(\Delta^{-1})_{kl}}{\alpha\gamma_{k-}}\\
\Delta_{lk}=-\frac{q_{c}^{(l)}(g_0^{-1})_{cd}q_{d}^{(k)}}{\alpha^2(1-\gamma_{k-}\gamma_{l-})} \\
M_{k} = \Psi_0(\gamma_{k-},\varsigma,\eta)
\end{eqnarray}
It may be verified directly, by expressing in terms of hyperbolic functions, that (\ref{eq:chiinv}), with $S_k$ given by(\ref{eq:Sk}) provides an inverse of $\chi$ for the particular background considered here and with 
\begin{equation}
\label{eq:mq}
m_{01}^{(k)}q_{01}^{(k^{'})}=m_{02}^{(k)}q_{02}^{(k^{'})}
\end{equation}

In the present we will use the background solution corresponding, among other cases, to various diagonal solutions of cosmological type and cylindrically symmetric gravitational waves \cite{BelinskyVerdaguer}
\begin{eqnarray}
 \label{eq:g0}
 (g_{0})_{11} =  \alpha e^{u_{0}}   \\
 (g_{0})_{22} =  \alpha e^{-u_{0}}  \\
 (g_{0})_{12} =  (g_{0})_{21} =0.
\end{eqnarray}
From now on we will fix our attention to cylindrically symmetric gravitational waves in which case we may take $\alpha=\rho$, $u_{0}=t=\beta$ \cite{BelinskyVerdaguer}. The metric is (\ref{eq:metric}) with $(z, t, x^a)\rightarrow(t, \rho, ix^a)$ \cite{BelinskyVerdaguer}.
The corresponding solution of (\ref{eq:linear system}) is given by 
\begin{eqnarray}
 \label{eq:Psi0}
 (\Psi_{0})_{11} =  \alpha(\gamma^{2}+2\frac{\beta}{\alpha}\gamma+1)^{1/2} e^{\frac{1}{2}\alpha\gamma+\beta}  \\
 (\Psi_{0})_{22} =  \alpha(\gamma^{2}+2\frac{\beta}{\alpha}\gamma+1)^{1/2} e^{-\frac{1}{2}\alpha\gamma-\beta}   
\end{eqnarray}

\section{The Lie-algabra valued one form $A$, the Fundamental Poisson Bracket and the Transition Matrix}
\label{sec: spectral current}

Following \cite{KorotkinNicolai} we consider the 'spectral parameter current' $A(\gamma)$ given by 
\begin{equation}
 \label{eq:current}
 A(\gamma) = \Psi_{,\gamma}\Psi^{-1}.
\end{equation}

From (\ref{eq:differentials}, \ref{eq:current}, \ref{eq:linear system}) at $\gamma = \pm 1$ we see how the definitions (\ref{eq:A(+-1)}) arise.

The dressing ansatz clearly leads to the following relation for $A(\gamma)$
\begin{equation}
\label{eq:A}
A(\gamma) = \chi_{,\gamma}\chi^{-1}+\chi A_{0} \chi^{-1}.
\end{equation}
where, 
\begin{equation}
\label{eq:A0inPsi}
A_{0} = \Psi_{0,\gamma}\Psi_{0}^{-1}
\end{equation}
and $R_k, S_k$ are given by (\ref{eq:Rk}), (\ref{eq:Sk}). It should be mentioned that there are no second order poles in $\gamma$ in (\ref{eq:A}) due to the condition $R_k \chi^{-1}(\gamma_k)=0$ which must be present by construction of the $R_k$ \cite{BelinskyVerdaguer}.  
From (\ref{eq:Psi0}) we obtain $A_{0}(\gamma)$: 
\begin{equation}
  \label{eq:A0ingamma}
  A_{0}(\varsigma,\eta,\gamma) = {\cal K}(\gamma)\, \mbox{I} + \frac{1}{2}\alpha \sigma_{3} = \frac{(\gamma+\frac{\beta}{\alpha})}{(\gamma^{2}+2\frac{\beta}{\alpha}\gamma+1)} \mbox{I} + \frac{1}{2}\alpha \sigma_{3},
\end{equation}  
where \(\sigma_{3}\) is the well known Pauli matrix
\begin{equation}
  \label{eq:sigma3}
  \sigma_{3} = \left(
     \begin{array}{ll} 
	    1  &  0 \\
      0  &\!\!\! -1 
      \end{array}
      \right) 
\end{equation}

Now we consider the pde's for $A(\gamma)$ wanting to obtain the fundamental Poisson bracket \cite{FaddeevTakhtajan} which has been obtained \cite{KorotkinNicolai} directly for $A(\gamma)$ considering first order poles of $A(\gamma)$. Of course the ultimate goal is possible quantization as in
\cite{Korepin, KorotkinNicolai}. The possible advantage of doing this in the BZ formalism is that it is a well understood technique with many solutions obtained over the years and possible quantization would open potentially the way for quantizing other backgrounds which correspond to various physical cases. Besides the quantization obtained in \cite{KorotkinNicolai} was not completed due to not fully developped representation theory of SL(2,R). 

Differentiating (\ref{eq:A}) using (\ref{eq: chi diff equation}) and the relations $[A(\gamma),\chi_{,\gamma}\chi^{-1} ] = -[A(\gamma), \chi A_{0} \chi^{-1}]$, \([A_{0}(\pm 1),A_{0}(\gamma)]=0\) and the identities 
$
(\gamma_{,\varsigma})_{,\gamma}=\frac{1}{\alpha(1-\gamma)^{2}}-\frac{1}{2\alpha}, \,\,\,\, (\gamma_{,\eta})_{,\gamma}=-\frac{1}{\alpha(1+\gamma)^{2}}+\frac{1}{2\alpha}
$
 we get:

\begin{eqnarray}
	\label{eq:A diff eq1}
A_{,\varsigma}(\gamma)=\left[ A_{+}, A(\gamma) \right] + A_{+,\gamma},\\
	\label{eq:A diff eq2}
  A_{,\eta}(\gamma)= \left[ A_{-}, A(\gamma)\right] + A_{-,\gamma},
\end{eqnarray}
where 
\begin{equation}
A_{+} = \frac{A(1)}{\alpha(1-\gamma)}-\gamma_{,\varsigma}A(\gamma)\, , \,\,\,\,
A_{-} = \frac{A(-1)}{\alpha(1+\gamma)}-\gamma_{,\eta}A(\gamma) 
\end{equation}
where we see that the equations 'decouple' for the two variables $\varsigma$ and $\eta$ a fact first observed in \cite{KorotkinNicolai-PRL}. These equations are evidently analogous to those in Chapter 6 of \cite{SamtlebenThesis}. The form of the background solutions considered here corresponds to a wide variety of seed solutions \cite{BelinskyVerdaguer} allowing one to reach a wide class of the solution space via the dressing ansatz.

The fundamental Poisson bracket \cite{FaddeevTakhtajan} characterizes the formulation of various integrable systems  \cite{FaddeevTakhtajan}, including gravity \cite{KorotkinNicolai}, and plays a significant role in the quantization of these models \cite{Korepin, KorotkinNicolai}. So we want to evaluate the bracket: 
\begin{equation}
\label{eq:FPBdef}
\newsavebox{\DERIVBOXZLMnew}\savebox{\DERIVBOXZLMnew}[2.5em]{$\otimes\hspace{-.4em}\raisebox{-.8ex}{\scriptsize ,}\hspace{.5em}$}\newcommand{\DeriveszlmN}{\usebox{\DERIVBOXZLMnew}}
\left\{A(\gamma)\!\! \DeriveszlmN \!\!\! A(\mu)\right\}_{jk,mn} = \left\{A_{jm}(\gamma),A_{kn}(\mu)\right\}
\end{equation}.

To obtain the Poisson bracket we may view the residues of the poles of $A(\gamma)$ \begin{equation}
A(\gamma)=A_0+\sum_{j}\frac{A_{j}}{\gamma-\gamma_{j}}
\end{equation} as the main variables \cite{Jimboetal}, Chapter IV Part 2 of \cite{FaddeevTakhtajan} (see chapter 6 of \cite{SamtlebenThesis}). Then we have 
\begin{equation}
\label{eq:AjPB}
\left\{A_{i}^{A}, A_{j}^{B}\right\} = \delta_{ij} f^{AB}_{C} A^{C}_{i}
\end{equation}
that is, $A_j$ are assumed to belong to a Lie algebra. The term involving the $A_0$ term is not significant for the Poisson bracket as $A_0$ satisfies the equation \(\frac{d}{d\varsigma}A_0(\gamma)=\frac{A_0(1)}{\alpha(1-\gamma)^2}-(\gamma_{,\varsigma})_{,\gamma} A_0(\gamma)\) 
and the equivalent equation for $\eta$, as may be directly checked and as should be expected.

Then the fundamental Poisson bracket (in the terminology of \cite{FaddeevTakhtajan}) may be written as \cite{FaddeevTakhtajan, KorotkinNicolai, SamtlebenThesis}
\begin{equation}
\label{eq:FPBAgammaAmu}
\newsavebox{\DERIVBOXZLMnewAB}\savebox{\DERIVBOXZLMnewAB}[2.5em]{$\otimes\hspace{-.4em}\raisebox{-.8ex}{\scriptsize ,}\hspace{.5em}$}\newcommand{\DeriveszlmN}{\usebox{\DERIVBOXZLMnewAB}}
\left\{A(\gamma)\!\! \DeriveszlmN \!\!\! A(\mu)\right\} = \left[\Pi/(\gamma-\mu),\mbox{I}\otimes A(\gamma) + A(\mu)\otimes\mbox{I}\right]
\end{equation}
with $\Pi$, the permutation matrix in 4 dimensions satisfying
\begin{eqnarray}
  \label{eq:Pisigma}
  \Pi = \frac{1}{2}\left( I\otimes I + \sum_{a=1}^{3}\sigma_{a}\otimes\sigma_{a} \right)\\
  \label{eq:Pi}
  \Pi = \left(
              \begin{array}{llll}
              1 & 0 & 0 & 0 \\
              0 & 0 & 1 & 0 \\
              0 & 1 & 0 & 0 \\
              0 & 0 & 0 & 1
              \end{array}
              \right)
\end{eqnarray}
$\sigma_a$ the Pauli matrices.
We write 
\begin{eqnarray}
  \label{eq:A(1)}
  A(\pm 1) & = & \left| \chi_{,\gamma}\chi^{-1} \right|_{\gamma=\pm 1} +  \chi(\pm 1) A_{0}(\pm 1) \chi^{-1}(\pm 1) \\
  & = & \left| \chi_{,\gamma}\chi^{-1} \right|_{\gamma=\pm 1} + \frac{1}{2}(\pm \mbox{I} + \alpha \Omega_{\pm}),
\end{eqnarray} 
where\begin{equation}\Omega_{\pm } = \chi(\pm 1) \sigma_{3} \chi^{-1}(\pm 1).\end{equation} 
Also it can be shown from (\ref{eq:chichiT}) \cite{BelinskyZakharov1} that $A(\gamma)$ satisfies
\begin{equation}
  \label{eq:PsiPsiT}
  \Psi^{T}(\frac{1}{\gamma},\varsigma,\eta) g^{-1}(\varsigma,\eta) \Psi(\gamma,\varsigma,\eta) = g^{-1}(\varsigma,\eta),
\end{equation}
which upon differentiation gives \cite{KorotkinNicolai, SamtlebenThesis},
\begin{equation}
  \label{eq:AAT}
  \gamma^{2} A(\gamma) g = g A^{T}(1/\gamma).
\end{equation}
With $k=\ln f$ and the Lagrangian \cite{KorotkinNicolai, BreitenlohnerMaison}
\begin{equation}
   \label{eq:Lagrangian}
   {\cal L} = \alpha \left( k_{,\varsigma\eta}+\mbox{Tr}\left(\frac{A(1)A(-1)}{\alpha^{2}}\right) \right)
\end{equation}
the canonically conjugate variables are \cite{KorotkinNicolai} ($\varsigma$$\, ,k_{,\varsigma}$) and ($\eta$\, , $k_{,\eta}$ ) and the field variables $A_{j}$ \cite{Jimboetal}. 
It may be verified that a two-time structure in the sense of \cite{Jimboetal, Bernard, KorotkinNicolai, SamtlebenThesis} is present.

Also it can be seen from (\ref{eq:A(+-1)}), $A(\pm 1)$ satisfies \cite{KorotkinNicolai}
\begin{equation}
  \label{eq:A(+1)=-A(-1)}
  A(\varsigma,\eta,\gamma = 1) = -A(\eta,\varsigma,\gamma=-1) ,
\end{equation}
since $\alpha = \varsigma-\eta$. 
It may be further noticed that the transformation 
\begin{equation}
\label{eq:involution}
{\cal \tau}:(\varsigma, \eta, \gamma)\rightarrow(\eta, \varsigma, -\gamma)
\end{equation}
 which is equivalent, for the particular choice of $\alpha$ considered here, to $(\alpha, \beta, \gamma)\rightarrow(-\alpha, \beta, -\gamma)$ is an involution of $A_0(\gamma)$, that is it satisfies 
\begin{equation}
  \label{eq:A0(+-gamma)}
  A_0(\varsigma, \eta, \gamma) = -A_0( \eta, \varsigma,-\gamma) ,
\end{equation}
Moreover ${\cal \tau}:\Psi_0(\gamma)\rightarrow -\Psi_0(-\gamma)$, ${\cal \tau}:\chi(\gamma)\rightarrow\chi(\gamma)$ hence from (\ref{eq:A}) we have 
\begin{equation}
\label{eq:tau(A)}
{\cal \tau}:A(\gamma)\rightarrow -A(-\gamma)
\end{equation}
Now observing (\ref{eq:A diff eq1}), (\ref{eq:A diff eq2}) and (\ref{eq:gamma equation}) we see that the transformation ${\cal\tau}$ is an `involution' of the differential equations sending essentially the one to the other ie 
\begin{equation}
{\cal \tau}:(\frac{{\mbox d}}{{\mbox d}\varsigma}, \frac{{\mbox d}}{{\mbox d}\eta}, \gamma)\rightarrow(\frac{{\mbox d}}{{\mbox d}\eta}, \frac{{\mbox d}}{{\mbox d}\varsigma}, \gamma^{\prime}=-\gamma)
\end{equation}
. Having noticed (\ref{eq:A(+1)=-A(-1)}) we define the transition matrix ${\cal T}(\varsigma, \eta, \gamma)$ 
\begin{equation}
\label{eq:transition matrix}
{\cal T}(\varsigma, \eta, \xi, \gamma) \equiv A(\varsigma, \eta, \gamma) A^{-1}(\eta, \xi, -\gamma) = -A(\varsigma, \eta, \gamma) A^{-1}( \xi, \eta, \gamma)
\end{equation}
Considering ${\cal T}_{,\varsigma}$ we obtain
\begin{eqnarray}
\label{eq:transition matrix path.o.exp.}
{\cal T}(\varsigma, \eta, \xi, \gamma) = {\cal P} e^{\int_{\varsigma}^{\xi} U(\jmath, \eta, \gamma)  d\jmath} \\
\label{eq: U}
U = A_{,\varsigma}(\varsigma, \eta, \gamma)  A^{-1}(\varsigma, \eta, \gamma)
\end{eqnarray}
where ${\cal P}$ denotes path ordered exponential.
Further considering ${\cal T}_{,\eta}$ 
we obtain 
\begin{eqnarray}
\label{eq:transition matrix eta}
{\cal T}_{,\eta}(\varsigma, \eta, \xi, \gamma) = V(\varsigma, \eta, \gamma) {\cal T}(\varsigma, \eta, \xi, \gamma) - {\cal T}(\varsigma, \eta, \xi, \gamma) V(\xi, \eta, \gamma)\\
V(\varsigma, \eta, \gamma) = {A}_{,\eta}(\varsigma, \eta, \gamma)  A^{-1}(\varsigma, \eta, \gamma)
\end{eqnarray}
that is we have obtained a transition matrix analogous to the one that is very common in integrable pde's \cite{FaddeevTakhtajan, Bernard} with the null coordinate $\eta$ playing the role of time. It should be mentioned that such a matrix was lacking for the equations of gravity in the presence of two commuting Killing vectors in the spectral current formulation. Of course the roles of $\eta$ and $\varsigma$ could be reversed with a different definition of ${\cal T}$ (that would be ${\cal T}^{'}=A(\varsigma, \eta, \gamma)A^{-1}(\varsigma, \vartheta, \gamma)$). The transition matrix is extensively used in integrable pde's and in their quantization \cite{FaddeevTakhtajan, Bernard, Korepin}. It should be stressed that the poles of $\chi$ and hence $A(\gamma)$ correspond essentially to the null trajectories of the solitons and are the light cones $w_k-\beta=\pm\alpha$ \cite{BelinskyVerdaguer}. So we have a way of relating the variables $A(\gamma)$ from ingoing light-cone to the outgoing light-cone and vice versa via the transition matrix ${\cal T}$. The transition matrix can be defined for general $\alpha$ since in that case the differential equation for gamma becomes \cite{SamtlebenThesis}
\begin{equation}
\label{eq:gamma equation general alpha}
\gamma_{,\varsigma}=\frac{\alpha_{,\varsigma}}{\alpha} \frac{\gamma\left(1+\gamma\right)}{\left(1-\gamma\right)},
 \gamma_{,\eta}=- \frac{\alpha_{\eta}}{\alpha} \frac{\gamma\left(1-\gamma\right)}{\left(1+\gamma\right)},
\end{equation}
and the factor \(\alpha_{,\varsigma}/\alpha\) has the same behaviour as \(\alpha\) when acted upon by \({\cal \tau}\).

\section{Conclusion and Discussion}
\label{sec:Conclusion}
We have obtained the fundamental Poisson bracket (\ref{eq:FPBdef}) for gravisolitons in the dressing formalism in equation (\ref{eq:FPBAgammaAmu}). 
Further a transition matrix (\ref{eq:transition matrix eta}) relating ingoing with outgoing null coordinates
was presented satisfying equations similar to the ones satisfied by the transition matrix for other integrable pde's.
 
 Now if we consider \(A(\pm 1)\) and  we can easily see that 
\begin{equation}
\label{eq:conn}
A(+1)=\alpha\Gamma^{a}_{\varsigma  b},A(-1)=\alpha\Gamma^{a}_{\eta b}
\end{equation}
where  \(\Gamma\)  are the well known Christoffel symbols. These two Christoffel symbols encode how the two-dimensional section \( {\mbox d}x_{a}/{\mbox d}\lambda\) of the `holonomy' \cite{Rovelli} changes when dragged along a geodesic curve in the $(\varsigma, \eta)$ plane. So they encode the geometry `within' the two-space $x^a$. The other two non-zero Christoffel symbols, involving indices $a,b$,  $\Gamma^{\eta}_{a b}$, $\Gamma^{\varsigma}_{a b}$ can be obtained from these two. If we want to obtain observables then in terms of (\ref{eq:conn}) a good candidate then is (\ref{eq:transition matrix}) as the total differential w.r.t. $\eta$, $\gamma$ derivative and $\eta$ derivative, of its trace vanishes for appropriate boundary conditions. So it should be possible to obtain integrals of motion that would involve the analogues of the classical relativity connection. Of course an analogous definition of${\cal T}$ for $\varsigma$ could be given in order to obtain integrals of motion for $\varsigma$. This might have a natural interpretation in the case of cylindrically symmetric gravitational waves where $\eta$ and $\varsigma$ have clear meaning as ingoing and outgoing coordinates.

\end{document}